\documentclass[aps,prl,twocolumn,superscriptaddress,nofootinbib,floatfix]{revtex4}

\usepackage{amsmath,amssymb,amsfonts,mathtools,bm}
\usepackage{graphicx}
\usepackage[colorlinks=true,linkcolor=blue,citecolor=blue,urlcolor=blue]{hyperref}

\newcommand{\ket}[1]{|#1\rangle}
\newcommand{\bra}[1]{\langle #1|}
\newcommand{\braket}[2]{\langle #1|#2\rangle}
\newcommand{\Ent}{\operatorname{Ent}}

\newcommand{\Kcyc}{\mathcal{K}}

\newcommand{\ML}{\mathrm{ML}}

\begin{document}

\title{Exact Entanglement-Depth Speed Frontier for Complete Quantum Charging}
\author{Wenlong Sun}
\affiliation{Department of Physics, Yanbian University, Yanji 133001, China}
\affiliation{Institute of Quantum Science and Technology, Yanbian University, Yanji, Jilin 133002, China}

\author{Gang Lu}
\affiliation{Division of Foundational Teaching, Guangzhou College of Technology and Business, Guangzhou 510850, China}

\author{Yuanfeng Jin}
\thanks{yfkim@ybu.edu.cn}
\affiliation{Department of Physics, Yanbian University, Yanji 133001, China}
\affiliation{Institute of Quantum Science and Technology, Yanbian University, Yanji, Jilin 133002, China}

\begin{abstract}
Complete quantum charging provides a sharp setting in which to ask how much multipartite entanglement is forced by speed itself.
For a closed \(N\)-qubit battery evolving from \(\ket{\downarrow}^{\otimes N}\) to \(\ket{\uparrow}^{\otimes N}\) under a time-independent Hamiltonian, we exactly solve the pure-state depth-constrained speed problem.
If the realized trajectory has entanglement depth at most \(k\), then the largest possible QSL-normalized rate \(\eta=\tau_{\rm QSL}/T\) is
\(\eta_{\max}(k)=\lceil N/k\rceil^{-1/2}\).
Conversely, an observed rate \(\eta\) certifies trajectory entanglement depth at least
\(\bigl\lceil N/\lfloor \eta^{-2}\rfloor\bigr\rceil\).
The mechanism is block orthogonalization: under a fixed product partition, complete charging forces all blocks to orthogonalize simultaneously, and the quantum speed limit converts this counting constraint into the speed bound.
Balanced cluster-flip evolutions saturate the bound, establishing an exact integer staircase frontier.
Thus fast complete charging cannot be explained by many small independently charging blocks; in particular, crossing the threshold \(\eta>1/\sqrt2\) certifies, for \(N>1\), the generation of genuine \(N\)-partite entanglement.
\end{abstract}

\maketitle

\begingroup
\renewcommand{\thefootnote}{\fnsymbol{footnote}}
\footnotetext[1]{\href{mailto:yfkim@ybu.edu.cn}{yfkim@ybu.edu.cn}}
\endgroup
\emph{Introduction.---}
A quantum battery turns work storage into a problem of controlled many-body
motion.  In a complete charge, an initially empty state is not merely raised in
energy; it is driven to an orthogonal many-body state.  The time required for
this motion is therefore set not only by the available energy scale, but also by
how many cells participate coherently along the actual state trajectory.

Since the early proposals of quantum batteries, collective charging has been
viewed as a possible route to an advantage over parallel local charging
\cite{Alicki2013,Binder2015,Campaioli2017,Ferraro2018}.  This viewpoint has led
to a broad literature on interacting many-body chargers and charging advantage
\cite{Andolina2019,Rossini2020,Gyhm2022,Andolina2025,Rinaldi2025},
open-system and dissipative protocols \cite{Pokhrel2025}, and experimental or
superabsorptive realizations \cite{Quach2022,Li2025}; see also the review
\cite{Campaioli2024}.  What is still missing is a model-independent way to read
the many-body resource of a fast complete charge directly from the realized
trajectory.

Collective charging alone does not specify this resource.
A Hamiltonian may contain highly nonlocal terms even when the state follows an
orbit that factorizes into small independent sectors.  Conversely, an apparently
simple two-level motion may encode the coherent reversal of a large many-body
domain.  The relevant resource should therefore be identified from the
trajectory, rather than from unused terms in a Hamiltonian expansion.  This
distinction is essential: entanglement is not a universal monotone of battery
performance, and coherence, correlations, control constraints, extractable work,
and dissipation can enter in different ways in different charging tasks
\cite{Hovhannisyan2013,Kamin2020,GyhmFischer2024,Campaioli2024}.  In this Letter
we isolate a setting in which the trajectory-level resource admits an exact
characterization: coherent complete charging.

We consider
\(H_B=(\omega/2)\sum_{j=1}^N(1+\sigma_j^z)\), whose empty and fully charged
states are \(\ket{\downarrow}^{\otimes N}\) and
\(\ket{\uparrow}^{\otimes N}\).  The transition
\(\ket{\downarrow}^{\otimes N}\mapsto\ket{\uparrow}^{\otimes N}\)
stores the full energy \(N\omega\) and is simultaneously an exact
orthogonalization task.  If the cells are flipped independently, the process is
made of \(N\) single-cell orthogonalizations, and the Fubini--Study speeds add
only in quadrature.  If all cells flip as one coherent block, the orbit becomes
a single collective orthogonalization and can attain the intrinsic quantum speed
limit.  Between these two limits lies a discrete trajectory invariant: the
number of independent block orthogonalizations that make up the complete charge.

This Letter determines the corresponding frontier exactly for finite-dimensional
pure-state complete charging generated by time-independent Hamiltonians.  If the
realized trajectory has entanglement depth at most \(k\), then
\begin{equation}
    \eta_{\max}(k)=\frac{1}{\sqrt{\lceil N/k\rceil}},
    \label{eq:intro_frontier}
\end{equation}
and therefore
\begin{equation}
    \Ent[U]\ge
    \left\lceil\frac{N}{\lfloor\eta^{-2}\rfloor}\right\rceil,
    \qquad \eta=\frac{\tau_{\rm QSL}}{T}.
    \label{eq:intro_witness}
\end{equation}
Here \(\Ent[U]\) denotes the maximum entanglement depth attained along the
realized trajectory.  The integer \(\lfloor\eta^{-2}\rfloor\) is the largest
number of independent block orthogonalizations compatible with the observed
charging rate.  Covering all \(N\) cells with no more than this number of blocks
forces at least one block to contain at least
\(\lceil N/\lfloor\eta^{-2}\rfloor\rceil\) cells, which gives
Eq.~\eqref{eq:intro_witness}.

A closely related recent work by Shi et al. proposed the observed-rate witness
\(\Ent[U]\ge \lceil N\eta^2\rceil\) and supported it in representative
complete-charging settings~\cite{Shi2025}.  Here we solve the corresponding
depth-constrained speed problem exactly under the present assumptions.  The
bound \(\lceil N\eta^2\rceil\) captures the coarse envelope, whereas
Eqs.~\eqref{eq:intro_frontier} and \eqref{eq:intro_witness} keep the integer
block count and give the exact staircase frontier.  In this form, the
speed--resource relation is determined by the realized orbit and its cyclic
support, not by unused Hamiltonian levels or by formal nonlocal terms that do
not take part in the motion.

\emph{Setting and theorem.---}
We consider a finite-dimensional \(N\)-qubit system evolving under a time-independent Hamiltonian,
\[
      \ket{\psi(t)}=e^{-iHt}\ket{\psi_0},
      \qquad
      \ket{\psi_0}=\ket{\downarrow}^{\otimes N} .
\]
Complete charging means that, for some time \(T\),
\(\ket{\psi(T)}=e^{i\phi}\ket{\uparrow}^{\otimes N}\), with
\(\braket{\downarrow}{\uparrow}=0\).
With the battery Hamiltonian introduced above, this endpoint condition stores the full energy \(N\omega\) and is an exact many-body orthogonalization.

Let
\[
      \Kcyc=\operatorname{span}\{e^{-iHt}\ket{\psi_0}:t\in\mathbb R\}
\]
be the cyclic subspace generated by the orbit, and set \(H_{\Kcyc}=H|_{\Kcyc}\).
We use the Mandelstam--Tamm/Margolus--Levitin quantum speed limit on this dynamical support~\cite{MandelstamTamm1945,MargolusLevitin1998,Anandan1990,Levitin2009,DeffnerCampbell2017},
\[
      \tau_{\rm QSL}
      =\max\left\{
      \frac{\pi}{2\Delta H},
      \frac{\pi}{2\langle H\rangle_{\ML}}
      \right\},
      \qquad
      \eta=\frac{\tau_{\rm QSL}}{T},
\]
where \(\Delta H\) is evaluated in the initial state and
\begin{equation}
      \langle H\rangle_{\ML}
      =\bra{\psi_0}\left(H_{\Kcyc}-E_{\min}^{\Kcyc}I_{\Kcyc}\right)\ket{\psi_0} .
      \label{eq:ML_energy}
\end{equation}
This cyclic-subspace convention removes spectator spectrum: energy levels outside \(\Kcyc\) are never populated and do not affect the orbit, the charging time, or the generated entanglement, although they would otherwise change the Margolus--Levitin energy offset.
We write
\[
      \tau_{\rm MT}=\frac{\pi}{2\Delta H},
      \qquad
      \tau_{\rm ML}=\frac{\pi}{2\langle H\rangle_{\ML}},
\]
so that \(\tau_{\rm QSL}=\max\{\tau_{\rm MT},\tau_{\rm ML}\}\).
Thus \(\eta\) is a property of the realized charging orbit.

For a pure \(N\)-partite state, the entanglement depth is at most \(k\) if the state is product across a partition whose blocks all have size at most \(k\)~\cite{Sorensen2001,Horodecki2009,Guhne2009,Hyllus2012,Toth2012}.
We denote the smallest such \(k\) by \(\Ent(\ket{\psi})\), and define the depth generated by the charging orbit as
\[
      \Ent[U]=\max_{0\le t\le T}\Ent(\ket{\psi(t)}) .
\]
A partition block may be internally entangled; it is required only to be product with respect to the other blocks.
Thus entanglement depth measures the size of the largest internally coherent block generated along the orbit.

The main result is the following depth-limited frontier.
If the complete-charging orbit satisfies \(\Ent[U]\le k\), then
\[
      \eta\le \frac{1}{\sqrt{\lceil N/k\rceil}} .
\]
The bound is attainable, and hence
\begin{equation}
      \eta_{\max}(k)=\frac{1}{\sqrt{\lceil N/k\rceil}},
      \qquad 1\le k\le N .
      \label{eq:depth_limited_frontier}
\end{equation}
Conversely, for any observed rate \(0<\eta\le1\), one certifies
\begin{equation}
      \Ent[U]
      \ge
      \left\lceil\frac{N}{\lfloor\eta^{-2}\rfloor}\right\rceil .
      \label{eq:staircase_theorem}
\end{equation}
Since \(\lfloor\eta^{-2}\rfloor\le\eta^{-2}\), Eq.~\eqref{eq:staircase_theorem} implies the smoother bound
\[
      \Ent[U]\ge \lceil N\eta^2\rceil .
\]
The difference is the integer count.
For a trajectory that is product across a fixed partition with \(m\) independent blocks, complete charging entails \(m\) block orthogonalizations; the quantum speed limit gives \(\eta\le1/\sqrt m\); hence a rate \(\eta\) allows at most \(\lfloor\eta^{-2}\rfloor\) such orthogonalizations.
Distributing \(N\) cells among no more than that many blocks gives Eq.~\eqref{eq:staircase_theorem}.

\begin{figure}[t]
\centering
\includegraphics[width=\columnwidth]{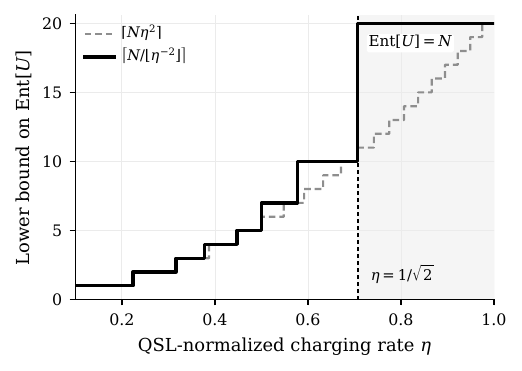}
\caption{
Exact integer speed--depth frontier for coherent complete charging, shown for \(N=20\).
The staircase gives the certified trajectory entanglement depth
\(D_{\rm cert}(\eta)=\lceil N/\lfloor \eta^{-2}\rfloor\rceil\)
from an observed QSL-normalized rate \(\eta=\tau_{\rm QSL}/T\).
The threshold \(\eta>1/\sqrt2\) certifies the generation of genuine \(N\)-partite entanglement.
}
\label{fig:bound_comparison}
\end{figure}

Figure~\ref{fig:bound_comparison} shows the effect of retaining the integer block count.
Crossing the threshold \(1/\sqrt m\) changes the maximum number of independent block orthogonalizations compatible with the observed rate.
Since all \(N\) cells must be covered by those blocks, the certified depth changes in discrete jumps.
At the maximal rate \(\eta=1\), the orbit must attain \(N\)-partite entanglement depth.

\emph{Proof sketch.---}
Assume that the complete-charging orbit has entanglement depth at most \(k\).
The only subtle point is that a pointwise low-depth condition may, a priori, use different product partitions at different times.
Since there are only finitely many product partitions, the corresponding closed time sets cover the interval \([0,T]\).
By the Baire category argument, at least one of them contains a nonempty relatively open subinterval of \([0,T]\).
On that interval the homogeneous polynomial equations defining the corresponding product variety vanish.
By analyticity of the finite-dimensional time-independent orbit, these equations then vanish along the whole orbit.
The fixed-partition lemma proved in the Supplemental Material makes this argument precise and yields one product partition supporting the entire trajectory.
Thus the orbit can be written, with respect to this fixed partition, as
\[
      \ket{\psi(t)}
      =
      \bigotimes_{\alpha=1}^{m}
      \ket{\phi_\alpha(t)}_{B_\alpha},
      \qquad |B_\alpha|\le k .
\]
Since the blocks cover all \(N\) cells,
\[
      m\ge \left\lceil \frac{N}{k}\right\rceil .
\]

Complete charging fixes the endpoint rays of every block.
For each block,
\[
      \ket{\phi_\alpha(0)}
      =
      \ket{\downarrow}^{\otimes |B_\alpha|},
      \qquad
      \ket{\phi_\alpha(T)}
      =
      e^{i\theta_\alpha}
      \ket{\uparrow}^{\otimes |B_\alpha|},
\]
and hence
\[
      \braket{\phi_\alpha(0)}{\phi_\alpha(T)}=0 .
\]
Thus a depth-\(k\) complete-charging orbit contains \(m\) block orthogonalizations.

For a product trajectory, the Fubini--Study speed satisfies
\[
      v^2(t)=\sum_{\alpha=1}^{m}v_\alpha^2(t),
\]
where \(v(t)\) is the Fubini--Study speed of the full path and \(v_\alpha(t)\) is the Fubini--Study speed of the \(\alpha\)-th block path.
Since each block connects orthogonal rays,
\[
      \int_0^T v_\alpha(t)\,dt\ge \frac{\pi}{2}.
\]
Minkowski's integral inequality therefore gives
\begin{align*}
      L
      &=
      \int_0^T
      \left(\sum_{\alpha=1}^{m}v_\alpha^2(t)\right)^{1/2}dt
      \\
      &\ge
      \left[
      \sum_{\alpha=1}^{m}
      \left(\int_0^T v_\alpha(t)\,dt\right)^2
      \right]^{1/2}
      \ge
      \frac{\pi}{2}\sqrt m .
\end{align*}
For time-independent evolution, \(v(t)=\Delta H\) is constant, so \(L=\Delta H T\). Hence
\[
      \Delta H T\ge \frac{\pi}{2}\sqrt m .
\]
Equivalently,
\[
      \frac{\tau_{\rm MT}}{T}
      =
      \frac{\pi}{2\Delta H T}
      \le
      \frac1{\sqrt m} .
\]

The Margolus--Levitin branch is evaluated on the cyclic support defined in Eq.~\eqref{eq:ML_energy}.
As shown in the Supplemental Material, the same block count gives
\[
      \frac{\tau_{\rm ML}}{T}\le \frac1m\le \frac1{\sqrt m} .
\]
Consequently the full QSL-normalized rate satisfies
\[
      \eta
      =
      \frac{\tau_{\rm QSL}}{T}
      \le
      \frac1{\sqrt m} .
\]
Combining this with \(m\ge \lceil N/k\rceil\) yields
\[
      \eta\le \frac1{\sqrt{\lceil N/k\rceil}} .
\]

Sharpness is obtained by an explicit cluster-flip construction.
Partition the \(N\) cells into \(m=\lceil N/k\rceil\) blocks of size at most \(k\), and let each block undergo a two-level coherent flip between its all-down and all-up states saturating the MT branch.
The product of these cluster flips has entanglement depth at most \(k\) and achieves
\[
      \eta=\frac1{\sqrt m}
      =
      \frac1{\sqrt{\lceil N/k\rceil}} .
\]
This proves the frontier in Eq.~\eqref{eq:depth_limited_frontier}.

\emph{Saturating construction.---}
The tight examples can be written explicitly.
Fix \(m=1,\ldots,N\), and divide the cells into \(m\) balanced blocks \(B_\mu\) with sizes
\(s_\mu\in\{\lfloor N/m\rfloor,\lceil N/m\rceil\}\).
For each block, set
\(\ket{D_\mu}=\ket{\downarrow}^{\otimes s_\mu}\) and
\(\ket{U_\mu}=\ket{\uparrow}^{\otimes s_\mu}\), and define the block flip
\[
      X_\mu=
      \ket{U_\mu}\bra{D_\mu}
      +
      \ket{D_\mu}\bra{U_\mu}
\]
on the two-dimensional subspace spanned by \(\ket{D_\mu}\) and \(\ket{U_\mu}\), extended by zero on its orthogonal complement.
In the full \(N\)-qubit Hilbert space, \(X_\mu\) is understood as this block operator on \(B_\mu\), tensored with the identity on all other blocks.
Consider
\[
      H_m=g\sum_{\mu=1}^{m}X_\mu,
      \qquad
      g=\frac{\pi}{2T}.
\]
This construction is used only to prove sharpness, not as a hardware proposal.

Starting from
\(\ket{\psi_0}=\bigotimes_{\mu=1}^{m}\ket{D_\mu}\), the orbit is
\[
      \ket{\psi(t)}
      =
      \bigotimes_{\mu=1}^{m}
      \left[\cos(gt)\ket{D_\mu}-i\sin(gt)\ket{U_\mu}\right].
\]
At \(t=T\), it becomes
\[
      \ket{\psi(T)}=(-i)^m\ket{\uparrow}^{\otimes N}.
\]
Each block follows a coherent GHZ-type flip, while different blocks remain product with respect to one another.
Hence
\[
      \Ent[U]
      =\max_\mu s_\mu
      =
      \left\lceil\frac{N}{m}\right\rceil .
\]

On the cyclic support of this orbit,
\[
      \Delta H_m=g\sqrt m,
      \qquad
      \langle H_m\rangle_{\ML}=mg .
\]
The Mandelstam--Tamm branch is therefore the active QSL branch, and
\[
      \eta=\frac{\tau_{\rm QSL}}{T}=\frac1{\sqrt m}.
\]
Since \(\lfloor\eta^{-2}\rfloor=m\), the construction saturates
\[
      \Ent[U]
      =
      \left\lceil\frac{N}{\lfloor\eta^{-2}\rfloor}\right\rceil
\]
for every block number \(m=1,\ldots,N\).
The cases \(m=N\) and \(m=1\) recover, respectively, parallel single-cell charging with
\(\eta=1/\sqrt N\) and fully coherent \(N\)-cell charging with \(\eta=1\).

Choosing \(m=\lceil N/k\rceil\) gives the depth-constrained optimum.
All block sizes are then at most \(k\), the construction has \(\Ent[U]\le k\), and the achieved rate is
\[
      \eta=\frac1{\sqrt{\lceil N/k\rceil}}.
\]
Together with the upper bound, this proves Eq.~\eqref{eq:depth_limited_frontier}.

\emph{Operational meaning and scope.---}
The theorem gives an orbit-level diagnostic.  A QSL-normalized rate
\(\eta\) constrains the realized trajectory, not just the form of the
Hamiltonian: there can be at most \(\lfloor\eta^{-2}\rfloor\)
independently orthogonalizing charging blocks.  It follows that
\[
      \Ent[U]\ge
      \left\lceil\frac{N}{\lfloor\eta^{-2}\rfloor}\right\rceil .
\]
In this sense, sufficiently fast complete charging requires multipartite
entanglement along the evolution.

The same counting picture explains the thresholds.  Parallel
single-cell charging has \(\eta=1/\sqrt N\), the sharp separable benchmark.
Hence any rate \(\eta>1/\sqrt N\) rules out a fully separable charging orbit
and certifies \(\Ent[U]\ge2\).  More generally, crossing
\(\eta>1/\sqrt m\) rules out a description with \(m\) or more independently
orthogonalizing blocks.  The certified depth therefore increases in integer
steps as the charging time approaches the QSL.  In particular, for \(N>1\),
the threshold \(\eta>1/\sqrt2\) already certifies genuine \(N\)-partite
entanglement.  The staircase is thus a consequence of the integer number of
complete block orthogonalizations, rather than of a continuously varying
entanglement measure.

The assumptions specify the benchmark.  We consider finite-dimensional
pure-state complete charging generated by a closed time-independent
Hamiltonian, with the exact endpoint condition
\[
\ket{\downarrow}^{\otimes N}\mapsto\ket{\uparrow}^{\otimes N}.
\]
Within this setting, the frontier follows from QSL geometry, endpoint
orthogonality, and the partition structure of entanglement depth; it does not
depend on a particular interaction graph or coupling pattern.  Balanced
cluster-flip trajectories saturate the bound for every admissible depth, so
the frontier is exact rather than only a necessary bound.

Approximate charging, mixed states, open dynamics, time-dependent controls,
work extraction, and thermodynamic figures of merit are outside this benchmark
and may require different resource measures.  For exact complete charging,
the QSL-normalized speed certifies an integer lower bound on trajectory-level
multipartite entanglement depth.

\section*{Acknowledgments}
This work was supported by the Natural Science Foundation of Jilin Province
under Grant Nos.~YDZJ202501ZYTS584 and 20250102018JC, and by the Liaoning
Provincial Joint Science and Technology Program (Natural Science
Foundation--General Program) under Grant No.~2025-MSLH-541.

\section*{Data availability}
No external datasets were used or analyzed in this work. The numerical values
shown in Fig.~\ref{fig:bound_comparison} were obtained by direct evaluation of the
analytic staircase formula stated in the manuscript.

\end{document}